# Magneto-acoustic study near the quantum critical point of the frustrated quantum antiferromagnet $Cs_2CuCl_4$

P. T. Cong,[1,2,a)] L. Postulka,[2] B. Wolf,[2] N. van Well,[2,3] F. Ritter,[2] W. Assmus,[2] C. Krellner,[2] and M. Lang[2,b)]

[1]*Dresden High Magnetic Field Laboratory, Helmholtz-Zentrum Dresden-Rossendorf, D-01314 Dresden, Germany*

[2]*Physics Institute, Goethe University Frankfurt, D-60438 Frankfurt(M), Germany*

[3]*Laboratory for Neutron Scattering and Imaging, Paul Scherrer Institute, CH-5232 Villigen, Switzerland*

Magneto-acoustic investigations of the frustrated triangular-lattice antiferromagnet $Cs_2CuCl_4$ were performed for the longitudinal modes $c_{11}$ and $c_{33}$ in magnetic fields along the *a*-axis. The temperature dependence of the sound velocity at zero field shows a mild softening at low temperature and displays a small kink-like anomaly at $T_N$. Isothermal measurements at $T < T_N$ of the sound attenuation $\alpha$ reveal two closely-spaced features of different character on approaching the material's quantum-critical point (QCP) at $B_s \approx 8.5$ T for $B \parallel a$. The peak at slightly lower fields remains sharp down to the lowest temperature and can be attributed to the ordering temperature $T_N(B)$. The second anomaly, which is rounded and which becomes reduced in size upon cooling, is assigned to the material's spin-liquid properties preceding the long-range antiferromagnetic ordering with decreasing temperature. These two features merge upon cooling suggesting a coincidence at the QCP. The elastic constant at lowest temperatures of our experiment at 32 mK can be well described by a Landau free energy model with a very small magnetoelastic coupling constant $G/k_B \approx 2.8$ K. The applicability of this classical model indicates the existence of a small gap in the magnetic excitation spectrum which drives the system away from quantum criticality.

## I. INTRODUCTION

Frustrated quantum antiferromagnets have attracted enormous interest in experimental and theoretical studies due to the discoveries of novel ground states and unusual excitations.[1] A model system for such frustrated low-dimensional (low-D) quantum magnets are S = ½ spins coupled antiferromagnetically on a simple triangular lattice.[2,3] $Cs_2CuCl_4$ represents a good realization of such a layered triangular-lattice Heisenberg quantum antiferromagnet.[4] The frustration results from a moderate antiferromagnetic (afm) exchange coupling constant $J/k_B$ = 4.3 K along the in-plane *b*-axis and a second in-plane coupling $J'$ ~ $J/3$ along a diagonal bond in the *bc*-plane, cf. inset b) of Fig. 1. Further couplings in this material include a weak inter-plane interaction $J''$ ~ $J/20$, leading to spiral antiferromagnetic (afm) long-range order at $T_N$ = 0.62 K (cf. Fig. 1), and a small anisotropic Dzyaloshinskii-Moriya (DM) interaction $D$ ~ $J/20$ creating an effective easy-plane anisotropy in the *bc* plane.[5] $Cs_2CuCl_4$ has attracted considerable interest due to its spin-liquid properties and its field-induced quantum phase transition at $B_s \approx 8.5$ T  for $B \parallel a$.[4-6] The quantum-critical point (QCP) at $B_s$ separates long-range afm order at lower fields from a fully-

a)Electronic mail:  t.pham@hzdr.de

b)Electronic mail: michael.lang@physik.uni-frankfurt.de

polarized ferromagnetic high-field state, cf. Fig. 1. With these characteristics, $Cs_2CuCl_4$ is well-suited to study the interplay of geometric frustration and low-d magnetism and their influence on the thermodynamic properties around a QCP. Generally, it is expected that the proximity of the QCP affects the thermodynamic properties also at finite temperatures. A manifestation of it includes, for example, the peculiar field dependence of the magnetic entropy at temperatures above a QCP, giving rise to a large magnetocaloric effect.[7] As has been discussed in Refs. [7, 8], this effect can be used for a very efficient magnetic cooling which can be of high relevance for certain applications. Here we explore the effects of the spin-lattice interactions and the quantum-critical behavior in the frustrated quasi-2d quantum magnet $Cs_2CuCl_4$ by using ultrasonic experiments.

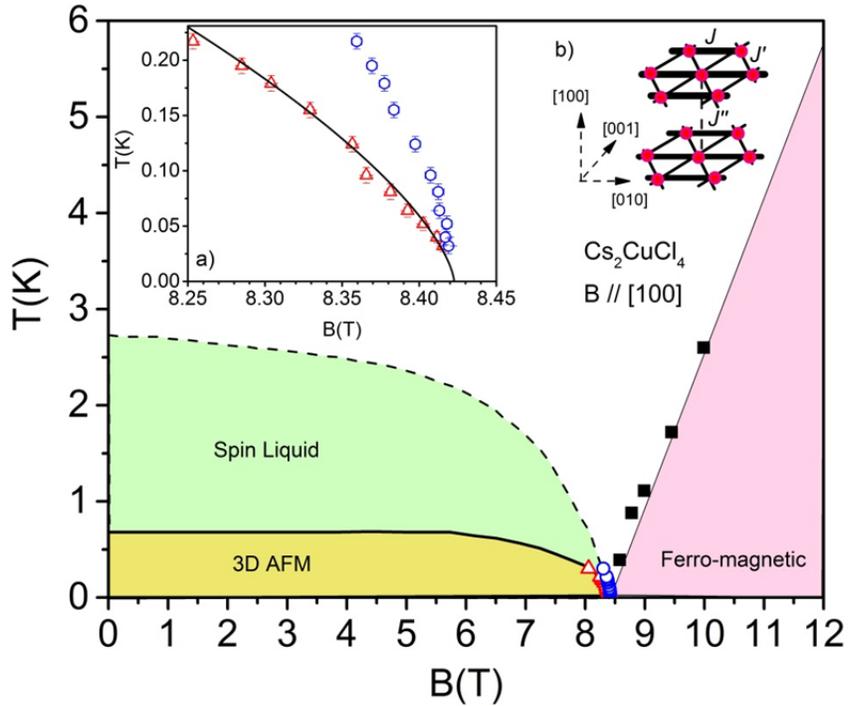

FIG. 1. Schematic phase diagram of $Cs_2CuCl_4$ as a function of temperature $T$ and an external magnetic field $B$ along the crystallographic $a$-axis (redrawn from Fig. 1 of Ref. [6]). The phase diagram includes: 3D long-range afm order ($T < T_N$) with spiral spin structure, a spin-liquid phase as well as a field-induced ferromagnetic ($B > B_c$) state. The spin excitations in the spin liquid phase have been shown[4] to propagate dominantly along the $b$-axis, implying the possibility to a dimensional reduction into a 1D scenario as proposed theoretically[2] and verified experimentally[19]. The thick solid line corresponds to a phase transition line whereas the dashed line and the thin solid line (connecting the open black squares taken from Ref. 13) indicate crossovers from a paramagnetic state (white area) to either the spin-liquid or the fully polarized (ferromagnetic) state. The red open triangles and the blue open circles indicate the position of anomalies in ultrasonic experiments discussed here. The inset a) gives details of the phase diagram in the proximity of the quantum-critical point on expanded scales. The inset b) shows the magnetic exchange paths in $Cs_2CuCl_4$. The $bc$ layer forms an anisotropic triangular lattice with $J'/J \sim 0.33$. The interlayer exchange coupling, indicated by the dashed line, amounts to $J'' \sim J/20$.

## II. EXPERIMENTAL DETAILS

Large and high-quality single crystals of $Cs_2CuCl_4$ were grown from aqueous solutions by an evaporation technique; see Ref. 9 for details. For the ultrasonic experiments, two parallel surfaces normal to the [100] and [001] directions were prepared. Piezoelectric polymer-foil transducers were glued to these surfaces for the generation and the detection of the ultrasound waves. The elastic properties were studied by measurements of the longitudinal sound waves (wave vector **k** ||

polarization **u**) propagating along the [100] and [001] directions, corresponding to the elastic modes $c_{11}$ and $c_{33}$, respectively. The elastic constant $c_{ij}$ is obtained from the sound velocity $v_{ij}$ by $c_{ij} = \rho v_{ij}^2$, where $\rho$ is the mass density. By using a pulse-echo method with a phase-sensitive detection technique[10] the relative change of the sound velocity $v_{ij}$ and the sound attenuation $\alpha$ were simultaneously measured as the function of temperature and magnetic field. A top-loading dilution refrigerator was used for measurements at 0.03 K < $T$ < 1.3 K and magnetic fields up to 12 T, whereas a $^4$He bath cryostat, equipped with a variable temperature insert, was employed for accessing temperatures $T$ > 1.3 K. The susceptibility data were recorded by using an ultra-high resolution ac-susceptometer adapted to a $^3$He–$^4$He dilution refrigerator. The samples, measured in the compensated-coil susceptometer, were prepared such that a nearly optimal filling factor was obtained. The amplitude of the applied ac-field was 0.5 mT at a frequency of 81 Hz.

### III. RESULTS AND DISCUSSION

$Cs_2CuCl_4$ exhibits a longitudinal elastic constant $c_{11} = 2.32 \cdot 10^{11}$ erg/cm$^3$ at 300 K which is approximately one order of magnitude smaller compared to other low-D magnets such as $SrCu_2(BO_3)_2$ indicating its high compressibility. Another peculiarity of this system includes the enormous increase in $c_{11}(T)$ of more than 20% upon cooling from room temperature down to 30 K, reflecting a very strong anharmonic elastic behavior in $Cs_2CuCl_4$ which has been confirmed in recent temperature-dependent X-ray studies.[11] As shown in the inset a) of Fig. 2, the $c_{11}(T)$ data at high temperatures, above about 20 K, can be well fitted by an empirical expression given by Varshni,[12] using the Debye temperature $\Theta_D/2 = 63$ K from specific heat experiments,[13] as a fixed parameter. From such fits to the high-temperature data, where the lattice contribution dominates and magnetic excitations are irrelevant, the elastic background can be extracted.

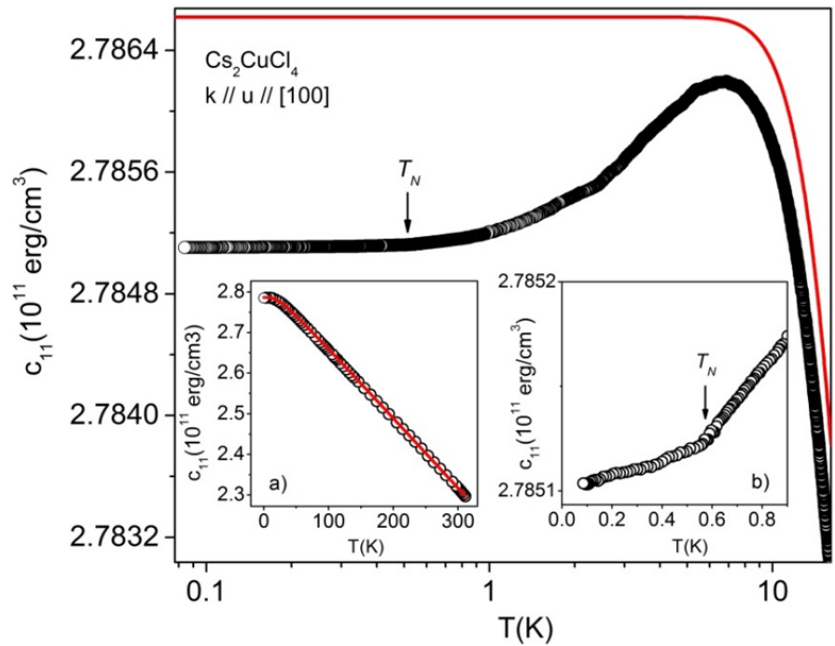

FIG. 2. The elastic constant for the acoustic $c_{11}$ mode (black open circles) as a function of temperature in zero field for 0.08 K $\leq T \leq$ 16 K in a semi-logarithmic representation. The red solid line represents the anharmonic elastic background which is obtained from fitting the experimental data according to an empirical expression discussed in Ref. 12. The fitting was performed for $T > 20$ K (corresponding to about $5J/k_B$, see inset a)), where the background contribution dominates, yielding the parameters presented in the text. Inset: a) $c_{11}(T)$ data (black open circles) and elastic background obtained from the fit (red line), b) $c_{11}(T)$ data around the phase transition to long-range afm order. The arrow indicates the kink-like anomaly at $T_N \sim$ 0.6 K.

The main panel of Fig. 2, where $c_{11}(T)$ in zero field is shown for 0.08 K $\leq T \leq$ 16 K on a semi-logarithmic scale, demonstrates that the deviations between the so-derived background elasticity and the experimental data are small ($\Delta c_{11}/c_{11}$ of order $10^{-3}$). These deviations can be assigned to the material's magnetic excitations and their coupling to the lattice via the magnetoelastic interaction. The experimental data reveal a mild softening of $c_{11}(T)$ below about 6 K. The long-range afm order at $T_N$ = 0.6 K manifests itself in a small kink-like anomaly at zero magnetic field (indicated by the arrow in the main panel as well as in the inset b) of Fig. 2). Compared to other low-d quantum magnets, especially coupled-dimer systems, such as $SrCu_2(BO_3)_2$ [14] or the natural mineral azurite[15], which exhibit elastic anomalies of a few percent due to a strong magnetoelastic coupling, the observed softening in $c_{11}(T)$ for $Cs_2CuCl_4$ of order $10^{-3}$ is rather small, indicating a weak magnetoelastic coupling in this case. Certain aspects of the elastic behavior of $Cs_2CuCl_4$ both as a function of temperature as well as magnetic field were already studied by several groups. This includes measurements of various elastic modes at high temperatures ($T > 100$ K) [16] as well as studies of the $c_{22}$ mode for fields applied along the $b$-axis.[17] In addition, the magneto-acoustic behavior of the $c_{22}$ mode in the spin-liquid region as well as in the ordered phase at low fields have been studied experimentally and theoretically.[18-19] It was found that the longitudinal modes $c_{11}$, $c_{22}$ and $c_{33}$ exhibit large linear changes when the material is cooled down from room temperature to liquid-nitrogen temperatures in zero field. [16] All longitudinal modes show similar anomalies in the ultrasonic attenuation $\Delta\alpha$ in the spin-liquid regime as well as in the afm ordered state.[17,19] Especially near the QCP, the longitudinal elastic constants $c_{xx}$ (xx = 11, 22, 33) and the corresponding attenuation exhibit a similar temperature and field dependence as shown in Ref. 21. From these experimental findings it is obvious that it is sufficient to investigate only one of the longitudinal modes to extract information about the underlying magnetoelastic coupling in $Cs_2CuCl_4$. Therefore we focus here on the $c_{33}$ mode and present results down to 32 mK for fields applied along the $a$-axis where the ordered phase is relatively simple, consisting only of the cone state.

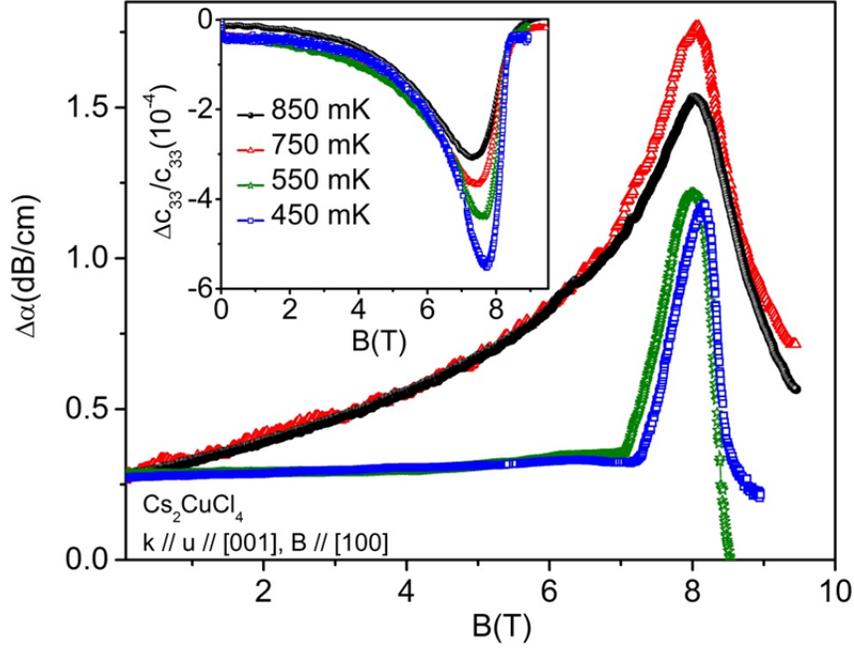

FIG. 3. Ultrasonic attenuation $\Delta\alpha$ of the $c_{33}$ mode as a function of magnetic field $B \parallel a$ at $T = 0.85$ K (black spheres) and 0.75 K (red triangles), i.e., above $T_N$, and at 0.55 K (green stars) and 0.45 K (blue (squares), corresponding to temperatures below $T_N$. Inset: Field dependence of the normalized $c_{33}$ mode at varying temperatures using the same symbol and color code as applied in the main panel.

Figure 3 displays the field dependence of the normalized $c_{33}$ mode (inset) together with the ultrasonic attenuation $\Delta\alpha$ (main panel) in the same symbol and color code at varying temperatures both above and below $T_N$. The attenuation curves for these temperatures exhibit a pronounced maximum around 8 T. We assign this effect to the coupling of the phonons to the magnetic excitations above the quantum-critical region. As has been shown by inelastic-neutron scattering experiments, these excitations are 2D in character with a dominant propagation along the $b$ direction[5, 6], suggesting a description within the proposed 1D dimensional-reduction scenario.[2] Whereas the position of the maximum stays practically unchanged around $B_s$ with increasing the temperature from 0.45 K up to 0.85 K, the shape of the anomaly alters significantly. For temperatures 0.75 K and 0.85 K, i.e., in the spin-liquid regime above $T_N$, the attenuation anomalies are broad and continuously increase up to the maximum around 8 T. This maximum in the attenuation decreases with increasing temperature and is visible up to at least 4 K (not shown). The presence of long-range afm order at low temperatures 0.45 K and 0.55 K alters the magnetic excitation spectrum, and with it, the field dependence of the ultrasonic attenuation $\Delta\alpha$. We find that within the afm ordered phase, $\Delta\alpha$ is essentially field independent up to around 6 T. Above 7 T $\Delta\alpha$ increases strongly and develops a pronounced maximum around 8 T. A clear anomaly in $\Delta\alpha$ corresponding to a magnetic phase transition at $B_N = 7.8$ T for 0.45 K and 7.2 T for 0.55 K cannot be discerned in the attenuation data. As was shown theoretically for $\Delta\alpha$ of the $c_{22}$ and the $c_{33}$ mode by a spin-wave expansion including the DM interactions, a field-independent attenuation is indeed expected for magnetic fields sufficiently below $B_s$.[18] The inset of Fig. 3 shows the $c_{33}$ elastic constant at the same temperatures at which the ultrasonic attenuation was measured. With increasing field this mode exhibits a softening followed by an upturn around $B_s$. The softening of the $c_{33}$ mode is similar for all data sets and fields below about 6 T. For higher fields, however, the curves split apart with a minimum that grows in size and slightly shifts to higher fields with decreasing temperature. It has been shown that the thermodynamic properties of $Cs_2CuCl_4$ can be described well by an afm Heisenberg chain model within a dimensional-reduction scenario.[19] By employing a 1D Ansatz and using a microscopic theory for the magnetoelastic interaction, it was found that the elastic behavior of the $c_{22}$ mode in the spin-liquid regime and the afm ordered phase can be well described up to about 6 T.[19] So far, no description based on a microscopic model is available for the magnetoelastic behavior around the QCP. As explicated in Ref. 20 such a dimensional-reduction scenario can also be employed to describe the inelastic neutron scattering data[4-6].

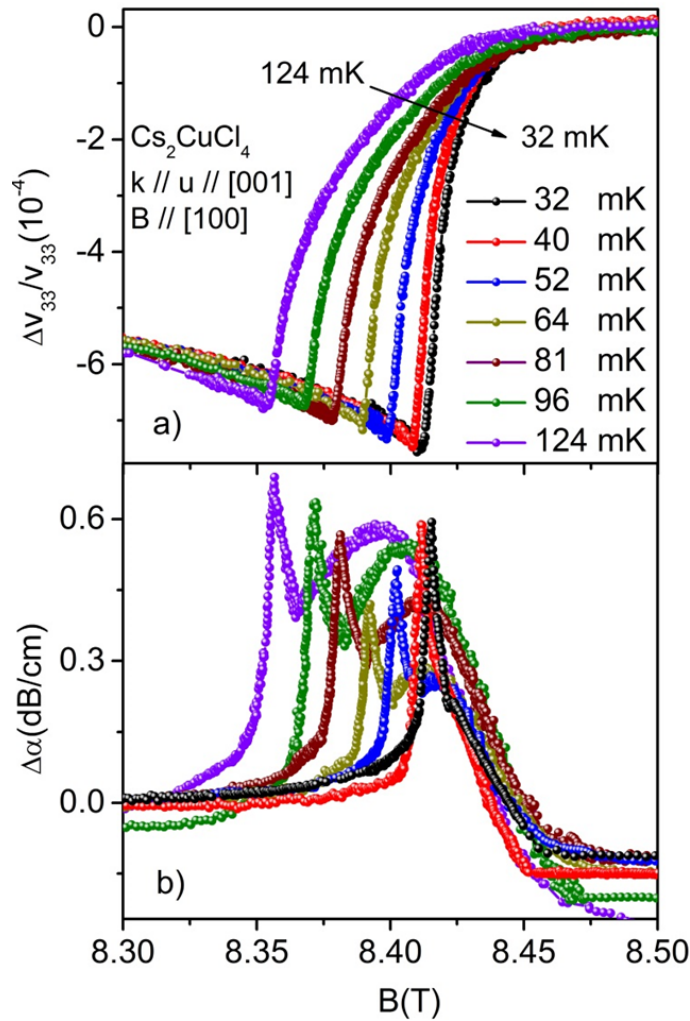

FIG. 4 a) Field dependence of the $c_{33}$ mode near the QCP for varying temperatures. b) Corresponding ultrasonic attenuation data using the same color code as in a).

Figure 4 shows the field dependence of the change of sound velocity (Fig. 4a) and ultrasonic attenuation (Fig. 4b) for the $c_{33}$ mode taken at different temperatures 0.032 K $\leq T \leq$ 0.13 K. Both data sets reveal pronounced anomalies around the QCP in a field range which narrows upon decreasing temperatures. Particularly striking are the double-peak features observed in the attenuation $\Delta\alpha$. They consist of a sharp anomaly on the low-field side which becomes more pronounced and is shifted to higher fields with decreasing temperature. This anomaly can be assigned to the phase transition into long-range afm order.[21] In addition, the data reveal a somewhat rounded maximum on the high-field side which significantly narrows in its width and becomes reduced in size upon cooling. We tentatively assign this rounded feature[21] to the crossover into the spin-liquid phase, identified by neutron scattering experiments[6], which coincides with peaks in the magnetic susceptibility[24] (broken line in Fig. 1) and specific heat.[13] At 32 mK - the lowest temperature of our $c_{33}$ measurements - the broad anomaly manifests itself only as a small shoulder on the high-field side of the sharp attenuation peak. The data suggest that both peaks merge together at the QCP as shown in the inset a) of fig. 1. Indications for two closely-spaced anomalies can also be revealed in the $c_{33}$ elastic mode, shown in Fig. 4a), by, e.g., looking at the derivative $\partial c_{33}/\partial B$ (not shown).[21] In Ref. 21 it was shown that the position of the steepest slope in the $c_{33}$ data corresponds to the sharp peak in $\Delta\alpha$ as expected for a second-order phase transition.[10] On the high-field side of the data in Fig. 4a) and 4b), i.e. for fields $B > B_s$, the system enters the fully-polarized state which is accompanied by the opening of a gap in the magnetic excitation spectrum for $B > B_s$.[6,13] As shown in Fig. 4a), this manifests itself in a bending which becomes increasingly sharp with decreasing temperature and a flattening of the elastic $c_{33}$ mode at higher fields.

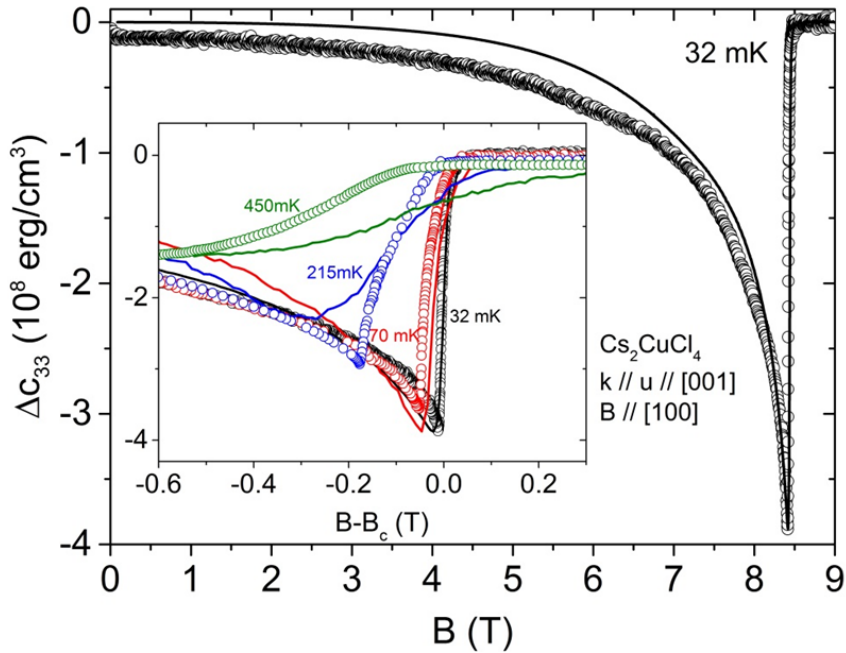

FIG. 5. The field dependence of $\Delta c_{33}$ (open circles) as a function of magnetic field near the QCP at 32 mK (main panel). The black solid line corresponds to calculations for $\Delta c_{33}$ based on a phenomenological description using experimentally determined susceptibility and magnetization data which are in accordance with literature results.[24] Inset: Data at varying higher temperatures on an extended B-scale relative to $B_c(T) = B_s(T)$.

For an estimate of the strength of the magnetoelastic coupling in this system we apply a Landau free energy model for the elastic constant based on an exchange-striction coupling.[21] As discussed in detail in Ref. 22 this Ansatz reveals a change of the elastic constant as a function of magnetic field of the form:

$$c_{33}(B) = c_{33}^0 - 4 \cdot G^2 \cdot n^2 \cdot \left(\frac{M(B)}{M_s}\right)^2 \cdot \left(\frac{\chi(B)}{M_s}\right). \quad (1)$$

Here $c_{33}^0$ is the background elastic constant, $M$ the magnetization, $\chi = \partial M/\partial B$ the magnetic susceptibility and $n$ the density of magnetic ions. Using experimentally determined magnetic susceptibility data, which are in accordance with literature results,[24] taken at the corresponding temperatures and the magnetization derived from integrating these $\chi(B)$ curves, allows for the modelling of the $c_{33}(B)$ data as shown in Fig. 5. The calculated curves shown in Fig. 5 are based on eq. (1) with the elastic background $c_{33}^0 = 2.78 \cdot 10^{11}$ and the density of magnetic ions $n = 4.32 \cdot 10^{21}$ cm$^{-3}$, in accordance with the structural data, and the magnetoelastic coupling constant $G = \partial J/\partial \varepsilon_L$ as the only adjustable parameter. Here $\varepsilon_L$ denotes the strain induced by the longitudinal sound waves propagating through the lattice which in turn modulates the distance between the magnetic Cu$^{2+}$ ions and, by this, their dominant exchange coupling $J$.[22] This Landau free energy model has been successfully applied to various low-D quantum spin systems, including the 2D S = ½ dimer system SrCu$_2$(BO$_3$)$_2$[15], the natural mineral azurite[14] and a Cu-coordination polymer[23] which is a good realization of an S = ½ afm Heisenberg chain. In the main panel of Fig. 5 we show the magnetic contribution $\Delta c_{33} = c_{33}(B) - c_{33}^0$ measured at the lowest temperature of 32 mK (black open circles) over the whole field range investigated. Also shown are the results of a fit (black solid line) to the data based on eq. (1) by using the magnetoelastic coupling constant G as a fit parameter. The figure shows that the model provides a good description of the data over the entire field range, including the region around $B_s$. From this fit we obtain a small magnetoelastic coupling constant G/$k_B$ ~ 2.8 K. It is somewhat smaller than G/$k_B$ ~ 20 K obtained for the quasi-1D Cu-coordination polymer studied in Ref. 22. However, it is about three orders of magnitude smaller than the coupling constant revealed for the 2D dimer system SrCu$_2$(BO$_3$)$_2$.[15] Given the dimensional reduction verified for Cs$_2$CuCl$_4$,[2] a magnetoelastic coupling constant similar in size to the one obtained for the Heisenberg spin-chain compound seems not surprising. We note, however, that at somewhat higher temperatures the fits based on eq. (1) yield more rounded features around $B_s$ which deviate from the $\Delta c_{33}$ data, cf. inset of Fig. 5. Given that the applicability of the phenomenological model implies the validity of Grüneisen scaling,[25] this is a surprising result in light of the generic divergence of the Grüneisen parameter (breakdown of Grüneisen scaling) upon approaching a QCP.[26] We speculate that the observation of classical behavior in the elastic constants at lowest temperatures is due to the presence of a small gap in the magnetic excitation spectrum already for $B \leq B_s$ which drives the system away from quantum criticality. Indeed, indications for such a gap due to the Dzyaloshinskii–Moriya interaction has been found in inelastic neutron-scattering experiments.[5]

## IV. CONCLUSIONS

We have investigated the magnetoelastic properties of $Cs_2CuCl_4$ for the longitudinal modes $c_{11}$ and $c_{33}$ in magnetic fields applied along the crystallographic *a*-axis. Acoustic anomalies have been found both in the temperature as well as the field dependence of these modes. The field-dependent measurements around $B_s$ display two distinct anomalies which are attributed to the transition into long-range afm order and signatures of the preceding spin-liquid state. These two anomalies in the sound attenuation $\Delta\alpha(B)$ merge at the QCP. The elastic data taken at the lowest temperature can be surprisingly well described by a classical phenomenological model with a small magnetoelastic coupling constant $G/k_B \sim 2.8$ K. The observation of rather classical behavior at lowest temperatures and the deviations from the classical behavior at somewhat higher temperatures are assigned to the opening of a small gap in the magnetic excitations spectrum for $B \leq B_s$ which drives the system away from quantum criticality.

## ACKNOWLEDGMENTS

We acknowledge financial support by the Deutsche Forschungsgemeinschaft via the SFB/TR49. PTC would like to thank the support of the HLD at HZDR, a member of the European Magnetic Field Laboratory (EMFL).

## REFERENCES

[1] S. Sachdev, Nature Physics **4**, 173 (2008) and ref. therein.
[2] L. Balents, Nature **464**, 199 (2010).
[3] H. C. Jiang, H. Yao, and L. Balents, Phys. Rev. B **86**, 024424 (2012).
[4] R. Coldea, D. A. Tennant, A. M. Tsvelik, and Z. Tylczynski, Phys. Rev. Lett. **86**, 1335 (2001).
[5] R. Coldea, D. A. Tennant, K. Habicht, P. Smeibidl, C. Wolters, and Z. Tylczynski, Phys. Rev. Lett. **88**, 137203 (2002).
[6] R. Coldea, D. A. Tennant, and Z. Tylczynski, Phys. Rev. B **68**, 134424 (2003).
[7] M. Lang, B. Wolf, A. Honecker, L. Balents, U. Tutsch, P. T. Cong, G. Hofmann, N. Krüger, F. Ritter, W. Assmus, A. Prokofiev, Phys. Status Solidi B **250**, 457 (2013).
[8] B. Wolf, A. Honecker, W. Hofstetter, U. Tutsch, M. Lang, Int. J. Mod. Phys. B **28**, 1430017 (2014).
[9] N. Krüger, S. Belz, F. Schossau, A. A. Haghighirad, P. T. Cong, B. Wolf, S. Gottlieb-Schoenmeyer, F. Ritter, and W. Assmus, Cryst. Growth Design **10**, 4456 (2010).
[10] B. Lüthi, G. Bruls, P. Thalmeier, B. Wolf, D. Finsterbusch, and I. Kouroudis, J. Low Temp. Phys. **95**, 257 (1994); B. Lüthi, *Physical Acoustics in the Solid State* (Springer, Berlin, 2005).
[11] N. van Well, K. Foyevtsova, S. Gottlieb-Schönmeyer, F. Ritter, R. S. Manna, B. Wolf, M. Meven, C. Pfleiderer, M. Lang, W. Assmus, R. Valentí, and C. Krellner, Phys. Rev. B **91**, 035124 (2015).

[12] Y. P. Varshni, Phys. Rev. B **2**, 3952 (1970).

[13] T. Radu, H. Wilhelm, V. Yushankhai, D. Kovrizhin, R. Coldea, Z. Tylczynski, T. Lühmann, F Steglich, Phys. Rev. Lett. **95,** 127202 (2005).

[14] P. T. Cong, B. Wolf, R. S. Manna, U. Tutsch, M. de Souza, A. Brühl, and M. Lang, Phys. Rev. B **89**, 174427 (2014).

[15] B. Wolf, S. Zherlitsyn, S. Schmidt, B. Lüthi, H. Kageyama, Y. Ueda, Phys. Rev. Lett. **86**, 4847 (2001).

[16] Z. Tylczynski, P. Piskunowicz , A. N. Nasyrov, A. D. Karaev, Kh. T. Shodiev, G. Gulamov, Phys. Status Solidi A **133,** 33 (1992).

[17] Sytcheva, O. Chiatti, J. Wosnitza, S. Zherlitsyn, A. A. Zvyagin, R. Coldea, and Z. Tylczynski, Phys. Rev. B **80**, 224414 (2009).

[18] A. Kreisel, P. Kopietz, P. T. Cong, B. Wolf, and M. Lang, Phys. Rev. B **84**, 024414 (2011).

[19] S. Streib, P. Kopietz, P. T. Cong, B. Wolf, M. Lang, N. van Well, F. Ritter, and W. Assmus, Phys. Rev. B **91**, 041108(R) (2015).

[20] M. Kohno, O. A. Starykh, L. Balents, Nature Physics **3**, 790 (2007).

[21] B. Wolf, P.T. Cong, N. Krüger, F. Ritter, M. Lang, J. of Phys.: Conference Series **400**, 032113 (2012).

[22] K. Kawasaki, Prog. Theor. Phys. **29**, 801 (1963).

[23] B. Wolf, S. Zherlitsyn, B. Lüthi, N. Harrison, U. Löw, V. Pashchenko, M. Lang, G. Margraf, H.-W. Lerner, E. Dahlmann, F. Ritter, W. Assmus, M. Wagner, Phys. Rev. B **69**, 092403 (2004).

[24] Y. Tokiwa, T. Radu, R. Coldea, H. Wilhelm, Z. Tylczynski, and F. Steglich, Phys. Rev. B **73**, 134414 (2006).

[25] E. Grüneisen, Ann. Phys. (Leipz.) **331**, 211 (1908).

[26] L. Zhu, M. Garst, A. Rosch, Q. Si, Phys. Rev. Lett. **91**, 066404 (2003).